\documentstyle[]{elsart}
\input{psfig.sty}
\begin{document}

\runauthor{L. Maraschi}

\begin{frontmatter}
\title{Correlated variability of Mkn 421 at  X-ray and TeV
wavelengths 
on timescales of hours.}
\author[]{L. Maraschi$^a$, G. Fossati$^b$, F. Tavecchio$^a$, L.
Chiappetti$^c$, 
A. Celotti$^b$,} 
\author[]{G. Ghisellini$^a$, P. Grandi$^d$, E. Pian$^e$, G.
Tagliaferri$^a$, 
A. Treves$^f$,}
\author[]{A.C.Breslin$^h$, J.H.Buckley$^i$,
D.A.Carter-Lewis$^j$, M.Catanese$^j$}
\author[]{ M.F.Cawley$^k$, D.J.Fegan$^h$, S.Fegan$^g$,
J.Finley$^l$, J.Gaidos$^l$, T.Hall$^l$} 
\author[]{A.M.Hillas$^m$, F.Krennrich$^g$, R.W. Lessard$^l$, C.Masterson$^h$} 
\author[]{P.Moriarty$^n$, J.Quinn$^h$, J.Rose$^m$, F.Samuelson$^j$, 
T.C.Weekes$^g$}
\address[]{Osservatorio Astronomico di Brera, Milano, Italy}
\address[]{SISSA/ISAS, Trieste, Italy}
\address[]{IFC/CNR, Milano, Italy}
\address[]{IAS/CNR, Roma, Italy} 
\address[]{TESRE/CNR, Bologna, Italy}
\address[]{Universita' dell'Insubria, Como, Italy}
\address[]{Whipple Observatory, AZ, USA} 
\address[]{UCD, Dublin, Ireland}
\address[]{Washington University, USA}
\address[]{Iowa State University, Ames, IA USA}
\address[]{NUI, Maynooth, Ireland}
\address[]{Purdue University, West Lafayette, IN USA}
\address[]{University of Leeds, Leeds, U.K.} 
\address[]{Galway-Mayo IT, Ireland}
\begin{abstract}
Mkn 421 was observed for about two days with BeppoSAX, prior to and
partly 
overlapping the start of a 1 week continuous exposure with ASCA in April 1998, 
as part of a world-wide multiwavelength campaign. A pronounced, well
defined, flare
observed in X-rays was also observed simultaneously at TeV energies
by the Whipple Observatory's 10 m gamma-ray telescope.
These data provide the first evidence that the 
X-ray and TeV intensities are well correlated on time-scales of
hours.
\end{abstract}


\end{frontmatter}

\section {Introduction}

It is now widely recognized that  the strong and variable non-
thermal emission
 observed in  BL Lac objects is due to a 
relativistic jet oriented at a small angle to the line of sight.
The  origin 
of such jets is not understood at present and observations of the
broad band
continuum offer a means to pin down radiation processes and derive
the
physical parameters of the jet including the bulk Lorentz factor of
the flow.

Mkn 421 is the brightest BL Lac object at X-ray and UV wavelengths
and the
first extragalactic source discovered at TeV energies (Buckley,
this workshop).
The emission up to X-rays is thought to be due to synchrotron
radiation from high energy electrons in the jet,
while it is likely that gamma-rays from GeV to TeV energies derive
from the same electrons
via inverse Compton scattering of soft photons  ( e.g., Ulrich,
Maraschi and Urry 1997,
and refs therein).

If the above model is correct, a change in the density and/or
spectrum
of the high energy electrons is expected to produce simultaneous
variations
at the frequencies emitted by the same electrons through the two
processes.
In particular, the two peaks present in the broad band spectral
energy
distribution (SED) should "correspond" to electrons of the same
energy.
Hence, a correlation between X-ray and TeV emission is expected.

Simultaneous
observations in the above two bands indeed  provided significant evidence 
of correlation
on relatively long timescales but due to insufficient sampling  did not
probe short timescales
(Catanese et al. 1997, Buckley et al. 1996).
A new campaign was organized  in 1998 to
obtain continuous coverage in X-rays for at least one week with ASCA
complemented by other space and ground based telescopes. 
Preliminary results are reported at this
conference by Takahashi and Urry.
 Here we discuss observations obtained between April 21 -- 24 
with the {\it Beppo}SAX satellite and the Whipple Cherenkov
telescope, preceding the start of the ASCA observations.

\section { Observations}

\subsection {{\it Beppo}SAX observations}

 The scientific payload
carried by {\it Beppo}SAX is fully described in Boella et al.
(1997a). 
The data of interest  here derive from three coaligned instruments,
the
Low Energy Concentrator Spectrometer
(LECS, 0.1-10 keV, Parmar et al. 1997), the Medium Energy
Concentrator Spectrometer
  (MECS, 2-10 keV, Boella et al. 1997b) and the Phoswich Detector
System (PDS, 12-300 keV, Frontera et al.
1997).

Observations with {\it Beppo}SAX were 
scheduled prior to but partially overlapping with the ASCA ones in
order to extend
the interval of time covered and to allow intercalibration.
The data reduction for the PDS was done using the XAS software
(Chiappetti
and Dal Fiume 1997), while for  the LECS and MECS  linearized
cleaned event 
files generated at the {\it Beppo}SAX Science Data Centre (SDC)
were used. No
appreciable difference was found extracting  the MECS data with the
XAS software.
Light curves were accumulated from each instrument 
with the usual choices for extraction radius and
background subtraction as described in Chiappetti et al. 1998.

\subsection {Whipple observations}

The observations of gamma-rays in the TeV energy bands were made
with the Whipple Collaboration's 10m Atmospheric Cherenkov
Telescope (Cawley et al., 1990; see also Buckley this Volume).

The Whipple camera utilizes fast circular-face photomultiplier
tubes arranged in a hexagonal pattern, with intertube spacing of
0.25 deg. In 1997 the camera was upgraded to have 331 pixels
giving a total field of view of 4.8 deg. The trigger condition for
this camera was that any two of the 331 phototubes register a
signal $>$ 40 photoelectrons within the coincidence overlap time of
8 nsec. Light-cones, which minimize the dead-space between the phototubes
and reduce the albedo effect, are normally used with this camera
but were not in place for the observations reported here. The
absence of the light-cones as well as the reduced reflectivity of
the mirrors (due to exposure to the elements) resulted in a
somewhat higher energy threshold than usual for these observations.

Observations were taken  on the nights of
April 21, 22, 23 and 24, 1998, prior
to the major observing campaign with ASCA which is reported
elsewhere (Takahashi this Volume). After the detection of a strong
flux on April 21 during an initial ON/OFF run ( the source is
tracked
 continuously for 28 minutes
and then an equivalent amount of time is spent observing a region
offset by 30 minutes in Right Ascension over the same range of
elevations and azimuth angles) the subsequent
observations were mostly taken in the TRACKING mode to maximize
coverage. In the latter  mode the source is tracked continuously
 and the background is estimated from the off-axis events.

To cover an observation period of 4 hours, observations were made
over a large range of zenith angles. Because the
collection area and energy threshold increase with zenith angle
the observed gamma-ray rates are convoluted with a changing
collection area and energy threshold. In addition the gamma-ray
selection is a function of zenith angle. Hence, the rate variations
are not linearly related to the observed flux variations of the
source. Thus it was necessary to determine the collection areas 
so that a light-curve with integral fluxes above a single energy
could be
derived. Unfortunately, contemporaneous observations of the Crab Nebula 
were not
available with sufficient statistics over all the zenith angles so
that this analysis relies entirely on Monte Carlo shower
simulations. 

The Monte Carlo code (ISU simulation package) was used to generate
gamma-ray induced showers at zenith angles of 20 deg. (45000
events), 45 deg. (25000 events) and at 55 deg. (25000 events)
zenith angles. The common energy range of all three collection areas covers the
range from 2 TeV up to tens of TeV. 
For comparing rate measurements at the three different
zenith angle regimes, it is necessary to set a common energy
threshold and to calculate the different collection areas.  This is
done here by normalizing the rate measurements at 45 and 55 deg.
to the collection area at z = 20 deg. and by setting a threshold
of 2 TeV, where the telescope has a good sensitivity at all three
zenith angle ranges.
 The results reported here are based on limited
statistics and are therefore preliminary. 
The aim of this analysis is to derive a normalized flux as a
function of time rather than absolute fluxes and energy spectra. 

\begin{figure}
\vskip -0.6 truecm
\hspace*{1.7cm}
\psfig{file=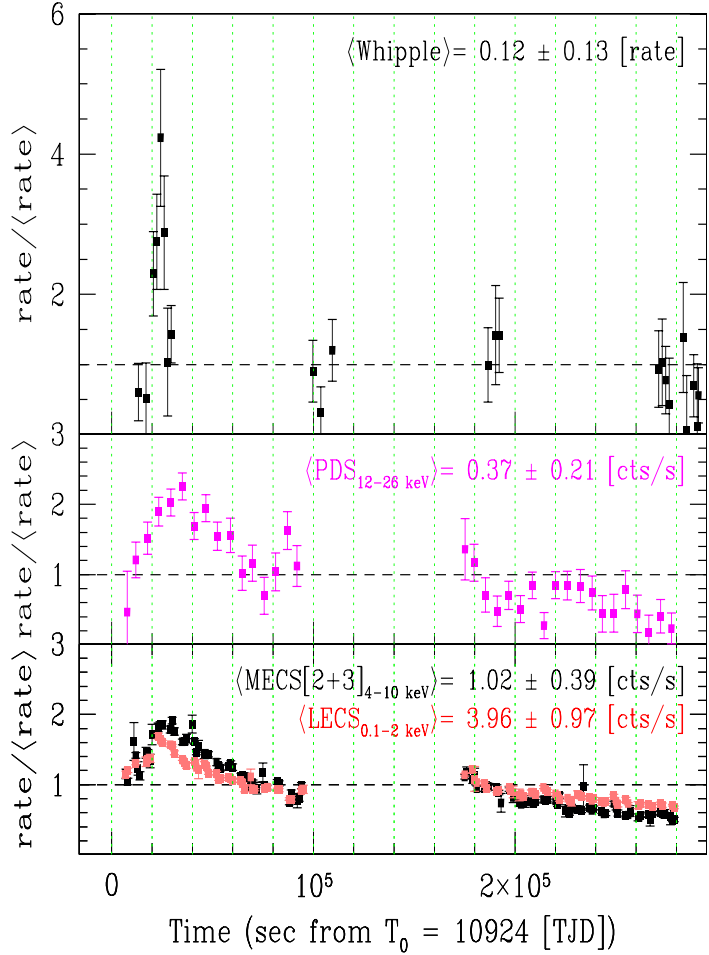,height=13.5cm}
\vskip -0.6 true cm
\caption{ Light curves of Mkn 421 at TeV and X-ray energies.
 The first panel shows the Whipple data ($>$2TeV), the second and third
 panel refer to {\it Beppo}SAX data with energy ranges as specified.
 All the count rates are normalized to their respective averages (given in each panel)
 over the observations shown.
}
\end{figure}

\section {Results}

A pronounced, well-defined, flare event is seen during the first
day of observation
with {\it Beppo}SAX.
The light curves in 3 energy bands (0.1-2 keV; 4-10 keV; 12-26 keV
 normalized to their respective means) are plotted in Fig. 1. 
The selected intervals were chosen so as to represent well-
separated
"effective" energies with reasonable statistics. Given the source
average 
spectrum and the instrumental response we can estimate them as
$\simeq 1$
keV for the LECS light curve in the 0.1 to 2 keV range, $\simeq 4$
keV for
the  MECS light curve in the 4-10 keV band and  15 keV for the PDS
light curve
 (15-26 keV).
Although the amplitude of the X-ray light curves increases with  energy
the peak to average intensity ratio is close to 2 in all X-ray bands (see 
Fig. 1). 
Further analysis is in progress with a view to quantifying possible
systematic differences in the shape of the light curves at
different energies.

The normalized event rates per 28 min interval from the Whipple
Cherenkov Telescope
 above the 2 TeV threshold for the four nights are also shown
in Fig 1, normalized to the average over the four nights.
 A clear peak is present with amplitude of a factor 4
and a halving time of about 1 hour. 
As discussed above the results are preliminary due to the use of
Monte Carlo
simulations to estimate
the telescope collecting area at different zenith angles.

The 0.1-2 keV, 4-10 keV and the 2 TeV peaks are simultaneous with
each other
within one hour but the halving time of the TeV light curve seems
definitely 
shorter than that of the LECS and MECS light curves. The 12-26 keV
light curve
seems to peak later but this is uncertain due to the limited
statistics.
The significance of possible leads/lags needs
further study. 

\section {Discussion and Conclusions}

The strong correlation between the TeV and X-ray flares on short
time-scales,
demonstrated by these data for the first time, supports models in
which the
high energy radiation arises from  the same population of high
energy
electrons that produce the X-ray flare via synchrotron radiation
and in particular from the same spatial region.
The most likely mechanism for the production of the
TeV photons is inverse Compton scattering of soft photons.
In particular, on the basis of the simultaneity of the peaks,
those electrons producing the 4-10 keV light curve
seem to be the best candidates for producing also the TeV flare.

The fact that the decay of the TeV flux is faster than that of the
keV flux
could imply that not only the high energy electron spectrum but
also
 the energy density of the target photons varies during the flare,
which could 
happen either in the synchrotron self-Compton  or in the mirror
scenario
(Ghisellini \& Maraschi 1996, Dermer, Sturner \& Schlickeiser 1997,
Ghisellini \& Madau 1996). Alternatively, electrons radiating 
at even higher synchrotron frequencies, whose light curve could not
be measured
with the present instrumentation, could have faster decay
timescales and
 be responsible for the TeV emission.

Although quantitative models are needed  in order to verify whether
any of
 the above scenarios can actually reproduce the observed
intensities and
timescales in the X-ray and TeV range, we anticipate that these
data 
will represent a very effective probe of the physical conditions
in the most active region of the Mrk 421 jet.

\end{document}